
\magnification=1200
\parskip=\medskipamount \overfullrule=0pt
\parindent=20truept
\font \titlefont=cmr10 scaled \magstep3
\font \namefont=cmr10 scaled \magstep1
\font \small=cmr8
\def\singlespace{\baselineskip=\normalbaselineskip}

\newcount\firstpageno \firstpageno=2
\footline={\ifnum\pageno<\firstpageno{\hfil}\else{\hfil
                                                  \rm\folio\hfil}\fi}
\def\frac#1/#2{\leavevmode\kern.1em
 \raise.5ex\hbox{\the\scriptfont0 #1}\kern-.1em
 /\kern-.15em\lower.25ex\hbox{\the\scriptfont0 #2}}
\def\degrees{\hbox{${}^\circ$\hskip-3pt .}}
\def\pp{\par\hangindent=.125truein \hangafter=1}
\def\aref#1;#2;#3;#4{\pp #1, {\it #2}, {\bf #3}, #4}
\def\abook#1;#2;#3{\pp #1, {\it #2}, #3}
\def\arep#1;#2;#3{\pp #1, #2, #3}

\def\Ob{\Omega_{\rm\scriptscriptstyle B}}
\def\Cth{C_{\rm th}}
\def\Cthij{C_{{\rm th,}ij}}
\def\mk{\,\mu{\rm K}}

\singlespace
\rightline{CfPA--93--TH--20}
\rightline{astro-ph/9308021}
\rightline{Revised version}
\rightline{August 1993}

\vskip 3pt plus 0.3fill
\centerline{\titlefont Are SP91 and COBE Inconsistent}
\vskip 15pt
\centerline{\titlefont with Cold Dark Matter?}

\vskip 3pt plus 0.3fill
\centerline{{\namefont Emory Bunn, Martin White,
Mark Srednicki}\footnote{${}^{\ast}$}{\small On leave from
Department of Physics, University of California, Santa Barbara, CA 93106.}
{\namefont \& Douglas Scott}}

\vskip 3pt plus 0.1fill
\centerline{Center for Particle Astrophysics}
\centerline{and Departments of Astronomy and Physics,}
\centerline{University of California, Berkeley, CA 94720}

\vskip 3pt plus 0.3fill

{\narrower
\centerline{ABSTRACT}
\baselineskip=15pt
We present results on the consistency of standard cold dark matter (CDM)
models with both the COBE normalization and the data from the 4th channel
of the 9-point South Pole scan.  We find that CDM models are consistent with
both experiments, a conclusion which is at odds with some other analyses.
This is partly due to our careful treatment of the temperature
autocorrelation function, but also derives from a realization that the
statistical conclusions depend strongly on assumptions about the prior
distribution for the normalization $Q$. }

\vskip 4pt

{\narrower
{\it Subject headings:} cosmic microwave background --- cosmology: theory
}

\vfill\eject
\baselineskip=16pt

\vskip0.2in
\vskip\parskip
{\bf 1. Introduction}
\vskip0.1in

There has been some confusion in the literature concerning the implications
for cold dark matter (CDM) models of the results of the COBE
(Smoot et al.~1992) and SP91 (Gaier et al.~1992) experiments.  In particular,
there have been contradictory claims (e.g.~Dodelson \& Jubas~1993,
G{\'o}rski et al.~1993) about the fundamental question: how incompatible are
standard CDM models with these anisotropy experiments?
Our intention is to provide a definitive answer to this question, giving the
details of our procedure.

We analyze in detail the data from the 4th (highest frequency)
channel of the SP91 9-point scan
(Gaier et al.~1992).  This data set seems most likely to represent actual
cosmic microwave background (CMB) fluctuations, and is the data set previously
analyzed by G{\'o}rski (1992), Dodelson \& Jubas (1993) and
G{\'o}rski et al.~(1993).
We wish to stress three points about our analysis:

(1)\ We use the theoretical two-point autocorrelation function
predicted for the SP91 measurement strategy by standard CDM models
($\Omega_0=1$, $h=0.5$, $n=1$, $0.01<\Ob<0.10$, $\Lambda=0$, $T/S=0$).
We notice that points whose angular separation is near the peak-to-peak
chopping angle have strong {\it negative} correlations.  This result,
explained below, is entirely due to chopping, and is completely independent
of the underlying cosmological model.
We compare the CDM autocorrelation function with the autocorrelation
function assumed by Gaier et al.~(1992) in their analysis.

(2)\ We quote results for an experiment-independent parameter, which we take
to be $Q\equiv Q_{\rm rms-PS}\equiv\left\langle Q^2_{\rm RMS}\right\rangle
{}^{\!0.5}$ (Smoot et al.~1992, Wright et al.~1993),
the r.m.s.~value of the quadrupole moment averaged over an
ensemble of universes.  In contrast, the usually quoted values of $C_0$ as a
function of $\theta_c$ (see below) must be disentangled from the observing
strategy to find information on the underlying spectrum of fluctuations.
Deconvolving theory from experiment is essential in obtaining a meaningful
comparison of experiments with different strategies measuring different parts
of the sky.  We remind the reader that the value of $Q$ measured by COBE
(Smoot et al.~1992, Wright et al.~1993) from the overall normalization of the
experimental two-point correlation function (assuming CDM with $n=1$ and no
gravity wave contribution) is $Q=17\pm5\mk$.
The quoted error includes the effects of cosmic and sample variance
(e.g.~White et al.~1993, Scott et al.~1993).

(3)\ We perform a maximum likelihood analysis, but investigate
sensitivity to the ``prior distribution''; that is, should we assume that,
a priori, equal intervals in $Q$ are equally likely (the usual assumption), or
equal intervals in $Q^2$ (which is simply proportional to the amplitude
of the power spectrum), or some other choice?
Different choices turn out to change the upper limit on $Q$ from SP91
(at various confidence levels) by tens of per cent.
Note that the choice of prior becomes less important in the limit that the data
have great ``discriminating power''.  At present, sensitivity to the prior
can be used as a measure of how constraining the 9-point scan is.

\vskip0.2in
\vskip\parskip
{\bf 2. The Temperature Autocorrelation Function}
\vskip0.1in

We begin with the theoretical two-point autocorrelation function for
CDM models.
Let $T(\theta,\phi)$ denote the temperature difference that the experiment
assigns to a point on the sky:
$$ T(\theta,\phi)=Q\,\sum_{\ell m}a_{\ell m}\, W_{\ell m}
\, Y_{\ell m}(\theta,\phi) \, , \eqno(1)$$
where the $W_{\ell m}$'s represent the window function of the experiment, and
the $a_{\ell m}$'s are random variables whose distribution must be specified
by a specific cosmological model.
In general, rotational invariance implies that
$$\langle  a_{\ell m} a^*_{\ell'm'}\rangle = C_\ell\,\delta_{\ell\ell'}
 \,\delta_{mm'}\, , \eqno(2)$$
where the angular brackets denote an ensemble average over the probability
distribution for the $a_{\ell m}$'s, and $C_\ell $ is normalized so that
$C_2=4\pi/5$.
For a pure Sachs--Wolfe, $n=1$ spectrum, $C_\ell^{-1}\propto \ell(\ell+1)$.
We compute the $C_\ell$'s for CDM models using power spectra provided by
Sugiyama (e.g.~Sugiyama \& Gouda~1992), which are essentially identical with
those computed by Bond \& Efstathiou (e.g.~1987).

The theoretical two-point autocorrelation function is given by
$$\eqalignno{
\Cthij &= Q^{-2}\,\left\langle T(\theta_i,\phi_i) T(\theta_j,\phi_j)
   \right\rangle \cr
\noalign{\medskip}
  &= \sum_{\ell m}C_\ell\,|W_{\ell m}|^2\,Y_{\ell m}(\theta_i,\phi_i)
\,Y^*_{\ell m}(\theta_j,\phi_j)\, , &(3)\cr}$$
where we have divided by $Q^2$ to make $\Cth$ dimensionless; there is then
no $Q$ dependence at all in $\Cth$.
Note that $C_{\rm th}$ is calculated using an underlying theory (e.g.~CDM)
together with the experimental window function, and does {\it not}
depend on any experimental data.
Thus $\Cth$ is not the same thing as the experimental correlation function
measured by the COBE team, which is computed by taking a sky average
of data without any theoretical input.

All the points in the 9-point scan of SP91 have the same zenith angle
but different azimuth.  In (3) we can write
$\theta_i = \theta = 27\degrees75$ and $\phi_i = \Delta\phi\times i$ with
$\Delta\phi\sin\theta = 2\degrees1$ being the angular spacing of points on the
sky.  Furthermore the chopping is done in the $\phi$ direction,
which implies that
$$W_{\ell m} = 2 H_0(m\alpha) \exp[-(\ell +{\frac1/2})^2\sigma^2/2] \eqno(4)$$
(Bond et al.~1991, Dodelson \& Jubas~1993, White et al.~1993),
where $H_0$ is the Struve function (Gradshteyn \& Ryzhik 1980),
$\alpha\sin\theta = 1\degrees5$ is half the peak-to-peak chopping angle on
the sky, and $\sigma=0.425\times1\degrees4$ is the Gaussian beam width of
the antenna.  Thus, equation~(3) becomes
$$ \Cthij = \sum_{\ell =2}^\infty C_\ell\,\exp[-(\ell+{\frac1/2})^2\sigma^2] \,
            \sum_{m=-\ell}^\ell 4 \, H^2_0(m\alpha) \,
            |Y_{\ell m}(\theta,0)|^2 \, \cos(m\phi_{ij})\, ,\eqno(5)$$
where $\phi_{ij} = \Delta\phi\times|i-j|$.  We find that $\Cth$ is
sensitive to the value of $\sigma$, and expect that departures from a
Gaussian beam profile may also significantly affect the results.

In Figure~1, we show $\Cth$ as a function of $\phi\sin\theta$, the separation
angle on the sky, for $\Ob=0.03$.  We see that it becomes strongly negative for
$\phi\sin\theta$ near $3^\circ$, corresponding to the peak-to-peak chopping
angle on the sky, $2\alpha\sin\theta$.  This is easy to understand:
the region halfway between the two negatively correlated points contributes
positively to one point and negatively to the other, which will clearly
produce an anticorrelation.

Results of small-scale CMB experiments are often quoted assuming an
underlying Gaussian autocorrelation function (GACF); this corresponds
to taking
$$Q^2 C_\ell = 2\pi\,\theta_c^2\,C_0\exp[-(\ell+{\frac1/2})^2\theta_c^2/2]\,,
                                                             \eqno(6)$$
where $\theta_c$ and $C_0$ are parameters.  In Figure~1, we show the
corresponding $\Cth$ as a function of $\phi\sin\theta$ for
$\theta_c=1\degrees2$ and $C_0/Q^2 = 7.9$.
The choice of $\theta_c=1\degrees2$ is the ``best-fit'' value
of Gaier et al.~(1992), which matches the peak of equation~(6)
to the peak of the SP91 window function in multipole space.
The choice of $C_0/Q^2 = 7.9$ results in $\Cth(0)=4.9$, the same
value as in the CDM model with $\Ob=0.03$.
Notice that for these particular choices of $\theta_c$ and $C_0$,
the anti-correlation at the nearest-neighbor separation of
$\Delta\phi\sin\theta=2\degrees1$ is well reproduced.
This agreement would not hold for general values of $\theta_c$ and $C_0$.

The signal-to-noise of the experiment is currently such that experimental
errors contribute large diagonal entries to the correlation matrix [see
equation~(8) below], reducing the effect of off-diagonal entries on the fit.
If the signal-to-noise were to improve significantly, then the off-diagonal
entries would play a larger role.
As an example, with the current data we find that a diagonal $\Cthij$
produces a very similar limit to that obtained in Gaier et al.~(1992), showing
that the assumed off-diagonal terms are playing little role in the analysis.
This can also be seen in Table~1, where the upper limits on $Q$ for a given
choice of prior distribution (see below) simply scale like $\Cth^{-1/2}(0)$
as $\Ob$ is varied.

\vskip0.2in
\vskip\parskip
{\bf 3. Maximum Likelihood Analysis}
\vskip0.1in
\nobreak

We now turn to the limits which can be placed on $Q$.
In accord with standard CDM models we consider underlying cosmological
fluctuations with a Gaussian probability distribution (not to be confused with
fluctuations having a Gaussian power spectrum, or GACF).
We assume that the experimental errors $\sigma_i$ for each $T_i$ are
uncorrelated and Gaussian distributed.
Then the unnormalized likelihood function for $Q$ is given by
$$ {\cal L}(Q) \propto {1\over\sqrt{\det K}}
           \exp\bigl[-{\frac1/2}\, T_i(K^{-1})_{ij} T_j\bigr]\, , \eqno(7)$$
where the matrix $K$ is
$$K_{ij}=Q^2\,\Cthij + \sigma_i^2\,\delta_{ij}\, . \eqno(8)$$

Equation~(7) assumes that the temperatures have no systematic errors.
However, SP91 has a possible, unknown, systematic offset and linear
gradient, and so their best-fit values have been removed from the measured
temperatures.  [In fact, as a consequence of the data binning procedure,
the published values of $T_i$ (Gaier et al.~1992) have a small residual
gradient which should be removed (Gaier~1993).] To project out a weighted
offset and linear gradient from the $T_i$, we write
$\widetilde{T}_a=R_{ai}T_i$ where $R$ is a $7\times9$ matrix
($a=1,\ldots,7$ and $i=1,\ldots,9$).  The seven rows of $R$ must be linearly
independent, and each row must be orthogonal to each of two nine-component
vectors:  one whose elements are $1/\sigma_i^2$, and one whose elements are
$(i-5)/\sigma_i^2$.  The choice of $R$ is otherwise arbitrary and does not
affect the results.  Then, instead of equation~(7), the unnormalized
likelihood function is given by
$$ {\cal L}(Q) \propto {1\over\sqrt{\det M}}
 \exp\bigl[-{\frac1/2}\,\widetilde{T}_a(M^{-1})_{ab}\widetilde{T}_b\bigr]\,
                                                          ,\eqno(9)$$
where $M_{ab}=R_{ai}K_{ij}R_{jb}^{\rm T}$.
Note that the procedure used by Bond et al.~(1991) is equivalent to removing an
{\it unweighted} offset and linear gradient from the data.

We can now use equation~(9) to set confidence levels on $Q$.
We must first choose a prior distribution for $Q$; that is, we must
decide whether equal intervals of $Q$, $Q^2$, or some other monotonic function
$f(Q)$, are equally likely a priori.
Given a choice of $f(Q)$, the upper limit on $Q$ at a confidence level
of $c$ is the solution $Q_{\rm max}$ to the equation
$$ c={\int_0^{Q_{\rm max}} {\cal L}(Q) \,df(Q) \over
      \int_0^\infty {\cal L}(Q) \,df(Q)}\, .     \eqno(10)$$
Due to the large measurement errors in the data set, assuming
$f(Q)=\log Q$, as recommended for an overall scale factor,
can be problematic (see e.g.~Readhead et al.~1989).  One is thus led to
consider prior distributions which are uniform in the scaling variable.
The usual choice (Bond et al.~1991, Dodelson \& Jubas~1993,
G{\'o}rski et al.~1993) is $f(Q)=Q$ (the ``bias'' parameter $b_{\rho}$
is proportional to $Q^{-1}$, so that $dQ = db_\rho/b^2_\rho$).
However, there is no compelling reason to assume a prior distribution
that is uniform in $Q$.  It is just as natural, for example, to assume that
the prior distribution is uniform in the power spectrum normalization $Q^2$,
or to make use of the COBE measurement of $Q$.
The choice of $f(Q)$ is important because it strongly affects the resulting
values of $Q_{\rm max}$.  Note that for ``good'' data, the choice of the
prior distribution should make very little difference, so ``prior dependence"
gives us a handle on the constraining power of the data.
This is illustrated in Figure~2, where we show the cumulative likelihood
($c$ vs.\ $Q_{\rm max}$) for four different choices of $\Ob$ and for
$f(Q)=Q$ and $f(Q)=Q^2$.
Clearly the effect of the prior distribution is significant.  This sensitivity
may be understood as being due to limited sky coverage (see Scott et al.~1993).

In Figure~2 we also show a band corresponding to the COBE
measurement of $Q=17\pm5\mk$ ($1\sigma$ errors).
We feel that this plot makes it clear that CDM models cannot be ruled out at
high levels of confidence from a combination of the COBE and SP91 9-point data.
This is also shown in Table~1, where we list the upper limits on $Q$ at the
95\% confidence level for the eight combinations of $\Ob$ and choice of prior
distribution $f(Q)=Q$ and $f(Q)=Q^2$.

\vskip0.2in
\vskip\parskip
{\bf 4. Conclusions}
\vskip0.1in

A reanalysis of the SP91 data from the 9-point scan shows no conflict with
standard
\eject\noindent
CDM models for any value of $\Ob$ and for $Q$ within the one-sigma
range specified by the COBE results, $Q=17\pm5\mk$, in contradiction
with the conclusions of some earlier analyses.
We find that the window function for SP91 is sensitive to the beam size
assumed and suspect that
deviations from a Gaussian beam pattern may significantly
affect the results.

\bigskip
We are extremely grateful to Naoshi Sugiyama for providing us with
CDM radiation power spectra.  We also thank Josh Gundersen, Phil Lubin,
Peter Meinhold, and especially Todd Gaier for helpful discussions.
E.B. acknowledges the support of an NSF graduate fellowship.
This work was supported in part by NSF Grant Nos.~PHY--91--16964
and AST--91--20005.

\vskip0.4in
\vskip\parskip

\centerline{
\vbox{ \offinterlineskip
\halign {\vrule#& \hfil#\hfil& \vrule#& \hfil#\hfil& \vrule#& \hfil#\hfil&
\vrule#& \hfil#\hfil& \vrule#\cr
\noalign{\hrule}
height2pt&\omit&&\omit&&\omit&&\omit& \cr
&\quad$\Ob$\quad&&\quad$C_{\rm th}(0)$\quad&& \ $Q_{\rm max}\,,\ dQ$\ &&
\ $Q_{\rm max}\,,\ d(Q^2)$\ &\cr
height2pt&\omit&&\omit&&\omit&&\omit&\cr \noalign{\hrule}
height2pt&\omit&&\omit&&\omit&&\omit&\cr
& 0.01&& 4.3&& 17&& 25&\cr
height4pt&\omit&&\omit&&\omit&&\omit&\cr
& 0.03&& 4.9&& 16&& 24&\cr
height4pt&\omit&&\omit&&\omit&&\omit&\cr
& 0.06&& 5.4&& 15&& 22&\cr
height4pt&\omit&&\omit&&\omit&&\omit&\cr
& 0.10&& 6.2&& 14&& 21&\cr
height2pt&\omit&&\omit&&\omit&&\omit&\cr \noalign{\hrule} }} }
\noindent Table 1: The 95\% confidence level upper limits,
$Q_{\rm max}\ (\mu$K), from the 9-point scan for CDM with
a range of $\Ob$ and for prior distributions $dQ$ and $d(Q^2)$.
Also listed is the value of $C_{\rm th}(0)$ for each $\Ob$.

\vskip0.4in
{\bf References}
\vskip0.1in
\frenchspacing
\parindent=0truept

\aref Bond, J. R., \& Efstathiou, G., 1987;MNRAS;226;655

\aref Bond, J. R., Efstathiou, G., Lubin, P. M. \& Meinhold, P. R., 1991;
Phys. Rev. Lett.;66;2179

\aref Dodelson, S. \& Jubas, J. M., 1993;Phys. Rev. Lett.;70;2224

\aref Gaier, T., Schuster, J., Gundersen, J. O., Koch, T., Meinhold, P. R.,
Seiffert, M. \& Lubin, P. M., 1992;ApJ;398;L1

\pp Gaier, T., 1993, private communication

\aref G{\'o}rski, K. M., 1992;ApJ;398;L5

\aref G{\'o}rski, K. M., Stompor, R. \& Juszkiewicz, R., 1993;ApJ;410;L1

\aref Readhead, A.C.S. et al.~1989;Ap.J.;346;566

\abook Scott, D., Srednicki, M., \& White, M., 1993;ApJ;submitted,
astro-ph/9305030

\aref Smoot, G.~F., et al., 1992;ApJ;396;L1

\aref Sugiyama, N. \& Gouda, N., 1992;Prog. Theor. Phys.;88;803

\abook White, M., Krauss, L., \& Silk, J., 1993;ApJ;in press,
astro-ph/9303009

\arep Wright, E. L., Smoot, G. F., Kogut, A., Hinshaw, G., Tenorio, L.,
Lineweaver, C., Bennett, C. L. \& Lubin, P. M., 1993;COBE preprint;No.~93-06

\nonfrenchspacing

\vskip0.2in
\vskip\parskip
{\bf Figure Captions}
\vskip0.1in
\parindent=0pt

Fig.~1.\ The theoretical temperature autocorrelation function,
$C_{\rm th}$, as a function of scan angle on the sky, $\phi\sin\theta$
($\phi$ is the azimuthal angle and $\theta$ the zenith angle).
The solid line is the autocorrelation function assuming the CDM
power spectrum for $\Ob=0.03$ and $h=0.5$.
The dashed line is the traditional Gaussian autocorrelation function (GACF)
approximation, with a correlation angle $\theta_c=1\degrees2$, processed
through the experimental observing strategy and matched to the CDM prediction
at $\phi=0$.  For comparison, the corresponding power spectra,
$\ell(\ell+1)C_\ell$, for both theories are shown in the inset.

Fig.~2.\ The cumulative likelihood as a function of power spectrum
normalization, $Q\ (\mu$K), for CDM models with $\Ob=0.01,0.03,0.06,0.10$
($\Ob$ increases to the left).  Curves are labelled by the assumed prior
distribution.  The hatched band is the $\pm1\sigma$ allowed range of
$Q$ from COBE.

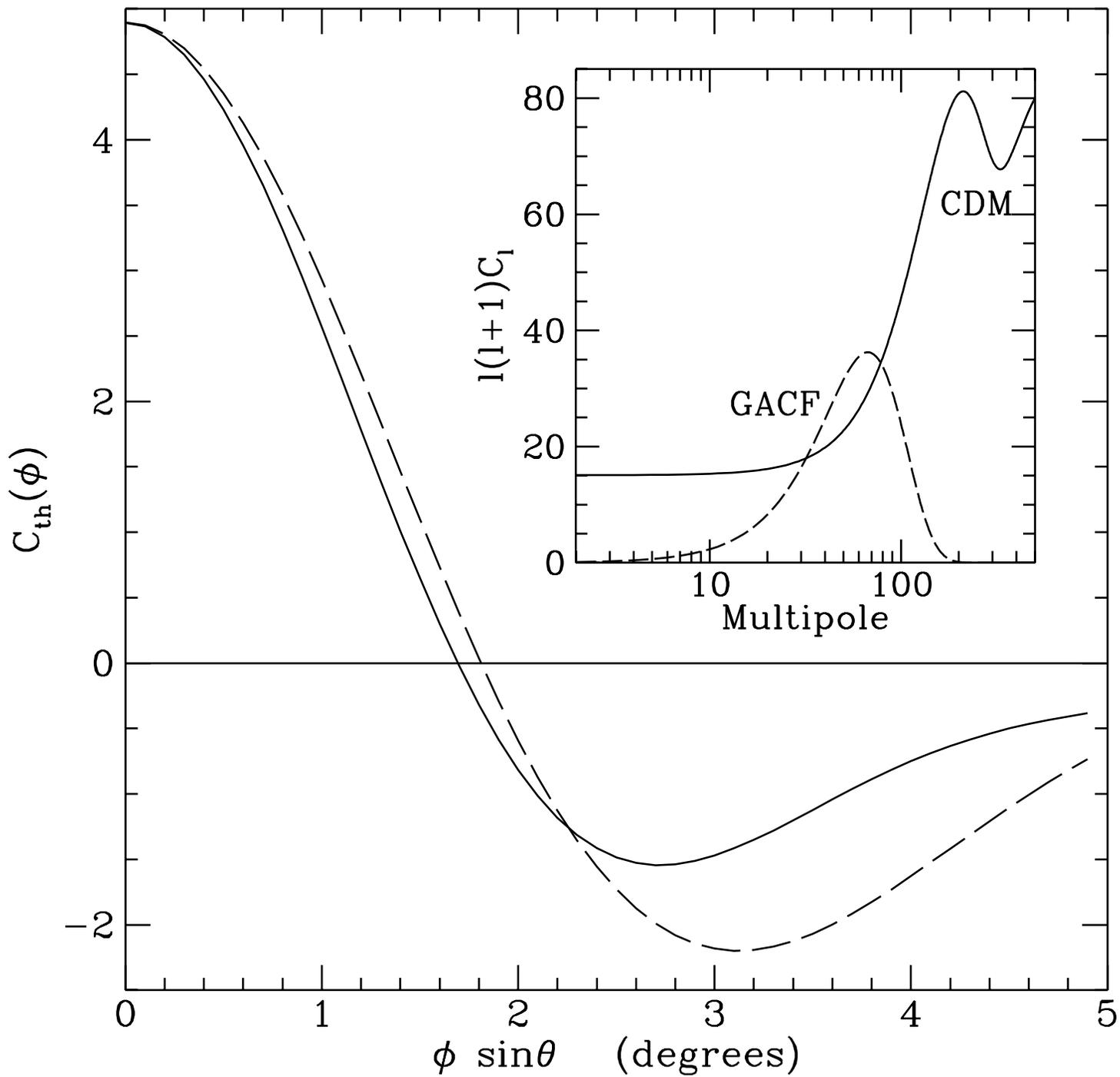
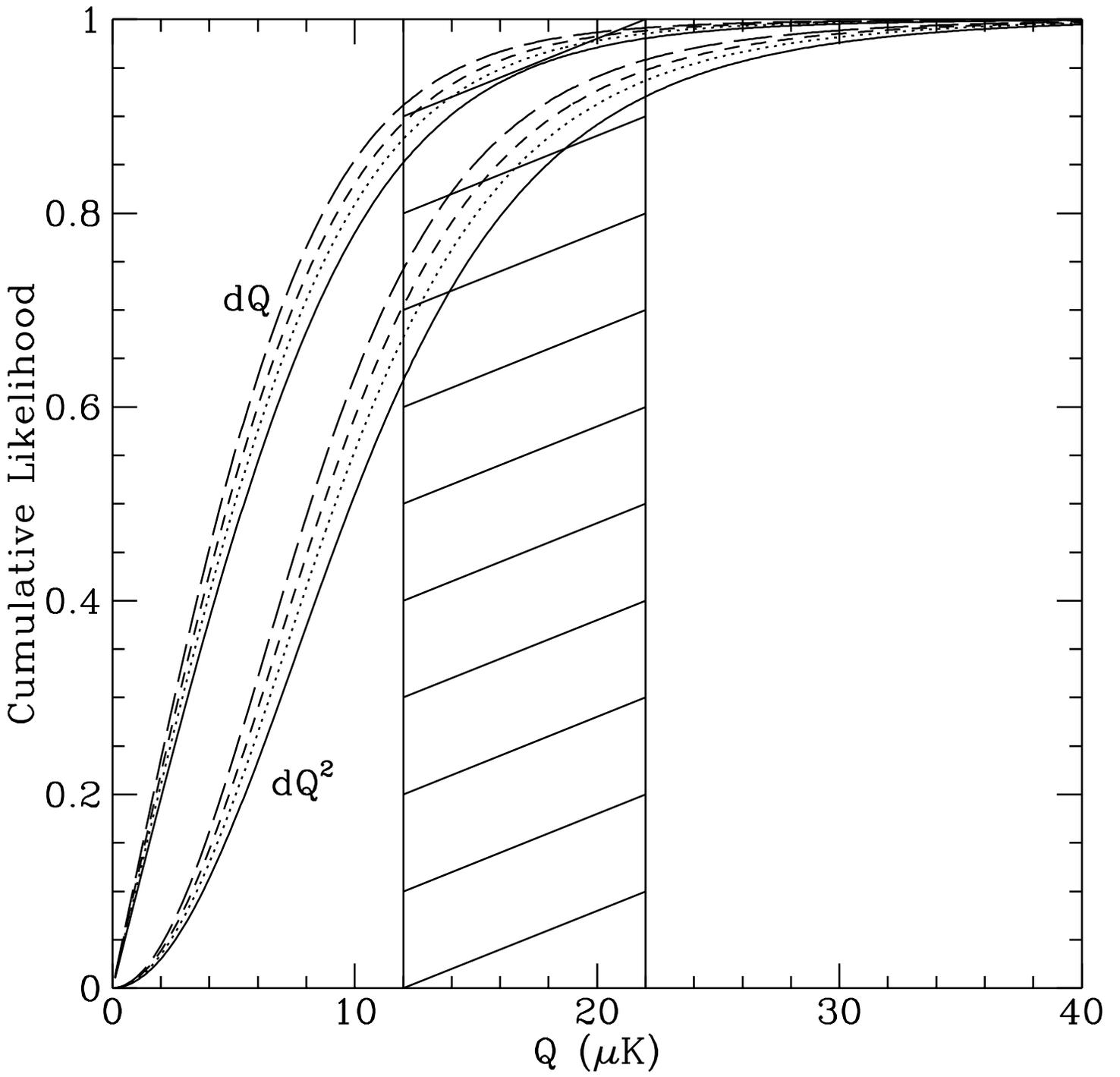
\end